\documentclass{article}
\usepackage{emulateapj}

\newcommand{\msun}{M$_{\odot}$}
\newcommand{\ergl}{ergs~s$^{-1}$}

\newcommand{\DTF}{{$D_{25}$}}
\newcommand{\mb}{$M_{B}$}
\newcommand{\cxo}{{\sl Chandra}}
\newcommand{\xmm}{{\sl XMM-Newton}}
\newcommand{\ros}{{\sl ROSAT}}
\newcommand{\etal}{et al.}
\slugcomment{Submitted to Astrophysical Journal}

\begin{document}

\title{Do Ultraluminous X-Ray Sources Exist in Dwarf Galaxies?}

\author{
Douglas~A.~Swartz\altaffilmark{1}
Roberto~Soria\altaffilmark{2}, and
Allyn~F.~Tennant\altaffilmark{3}}

\altaffiltext{1}{Universities Space Research Association,
    NASA Marshall Space Flight Center, VP62, Huntsville, AL, USA}
\altaffiltext{3}{Mullard Space Science Laboratory,
    University College London, Holmbury St. Mary, Surrey RH5 6NT, UK}
\altaffiltext{3}{Space Science Department,
    NASA Marshall Space Flight Center, VP62, Huntsville, AL, USA}

\begin{abstract}
A thorough search for Ultraluminous X-ray source candidates within
the Local Volume is made. 
The search spatially matches potential ULXs detected in X-ray images or 
cataloged in the literature with galaxies tabulated in the 
{\sl Catalog of Neighboring Galaxies} compiled by Karachentsev \etal\ (2004). 
The specific ULX frequency (occurrence rate per unit galaxy mass) is found to be a
decreasing function of host galaxy mass for host masses above 
$\sim$10$^{8.5}$~\msun. 
There is too little mass in galaxies below this point to determine  
if this trend continues to lower galaxy mass.
No ULXs have yet been detected in lower-mass galaxies.
Systematic differences between dwarf and giant galaxies that may explain
an abundance of ULXs in dwarf galaxies 
and what they may imply about the nature of ULXs are discussed.

\end{abstract}

\keywords{galaxies: general --- X-rays: galaxies --- X-rays: general}

\section{Introduction}  

Ultraluminous X-ray sources (ULXs) are defined as the 
most X-ray luminous, typically $L_{\rm X} \gtrsim 10^{39}$~\ergl,
non-nuclear point-like objects in nearby galaxies. 
What environments and what types of stellar systems give rise to the ULX
phenomenon is a subject of considerable topical interest.

The study of ULXs as a class has matured beyond mere number counts.
The general approach is to search for statistically-significant correlations 
between ULXs and properties of their galactic environments.
In this way, one hopes to eventually understand what physical processes 
are responsible for their formation, what influence they have on
their environment, and how they have evolved over cosmic time. 
Studies of ULXs as a class have reached some robust conclusions;
 ULXs are associated with active star formation, at least on galaxy-wide scales
(Swartz \etal\ 2004; Gilfanov \etal\ 2004; Liu \etal\ 2006; Winter \etal\ 2006)
  being more common and more luminous in starburst and interacting/merging galaxies
(e.g., Zezas \etal\ 2002, Gao \etal\ 2003) than in normal galaxies and 
occurring more frequently in the past (Lehmer \etal\ 2006) consistent with the
 observed rise in star formation density with redshift.
ULXs are less luminous in early-type galaxies, potentially revealing a secondary 
  population of unusually bright low-mass X-ray binaries
 or perhaps fortuitous beaming of a fraction
   of this population towards the observer 
(Irwin 2004; Swartz \etal\ 2004; Liu \etal\ 2006;
 see also King 2004).

There are circumstantial reasons to suspect that there may be differences 
between the rate of ULXs in nearby dwarf and giant galaxies.
Dwarf galaxies typically lack organized structure like bars and 
density waves that drive star formation yet their star formation rate (SFR) per
unit area can be comparable to those of spiral galaxies (Hunter \& Gallagher 1986).
Dwarf galaxies have evolved more slowly,
retaining a higher gas fraction than giants in the current epoch (Geha \etal\ 2006)
and resulting in a lower metallicity (Lee \etal\ 2006 and references therein).
Environmental factors, such as ram-pressure stripping, can strongly 
affect star formation in dwarfs
whereas star formation in giant field galaxies is determined more by
their merger history (Haines \etal\ 2007).
If ULXs (or the luminous subset of ULXs) are preferentially associated with 
 star formation, then there may be differences in the rates of ULXs in dwarfs
compared to giants because of the different internal and environmental 
influences on star formation activity in galaxies of different mass.

The ULX population in the Local Volume  ($D\le10$~Mpc) is investigated in the 
following. \S~2 describes the sample of galaxies included in the study and 
defines three subsamples with differing selection criteria and hence different
sample biases. The results of the search for ULXs in these subsamples are
presented in \S~3 where it is shown that the specific ULX frequency (number per
unit galaxy mass) increases towards lower galaxy mass down to a limit of 
$\sim$3$\times$10$^{8}$~\msun\ below which no ULXs have yet been detected.
The significance of this result is discussed in \S~4.

\section{The Samples}  

By definition, ULXs are the most luminous non-nuclear sources in 
galaxies. Nevertheless, their 0.5$-$8.0~keV flux, 
$f_{\rm X}>8.3\times10^{-12}/D^2$~erg~cm$^{-2}$~s$^{-1}$
(for a $L_{\rm X}=10^{39}$~\ergl\ source; the traditional definition of a ULX),
quickly falls below detection sensitivities of most X-ray observatories 
for sources beyond $D\sim20-30$~Mpc when observed for typical exposure times.
Similarly, properties of dwarf galaxies are best known for nearby objects.
In this case, the galaxies of interest 
are the least intrinsically luminous and have
the smallest physical size. For these reasons, a study of ULXs in dwarf galaxies 
is currently restricted to nearby objects within roughly the Local Volume. 
The recent compilation of galaxy properties
by Karachentsev \etal\ (2004, hereafter K04) is purported to be about 70-80\%
complete out to a distance $D=8$~Mpc. The catalog includes galaxies brighter than
$B_t \sim 17.5$~mag and angular sizes larger than \DTF$\sim$0.4\arcmin.
This Catalog of Neighboring Galaxies will be referred to as the CNG in this paper
and various samples of galaxies and their properties will be taken from the 
values reported in K04.

\begin{center}
\includegraphics[angle=-90,width=\columnwidth]{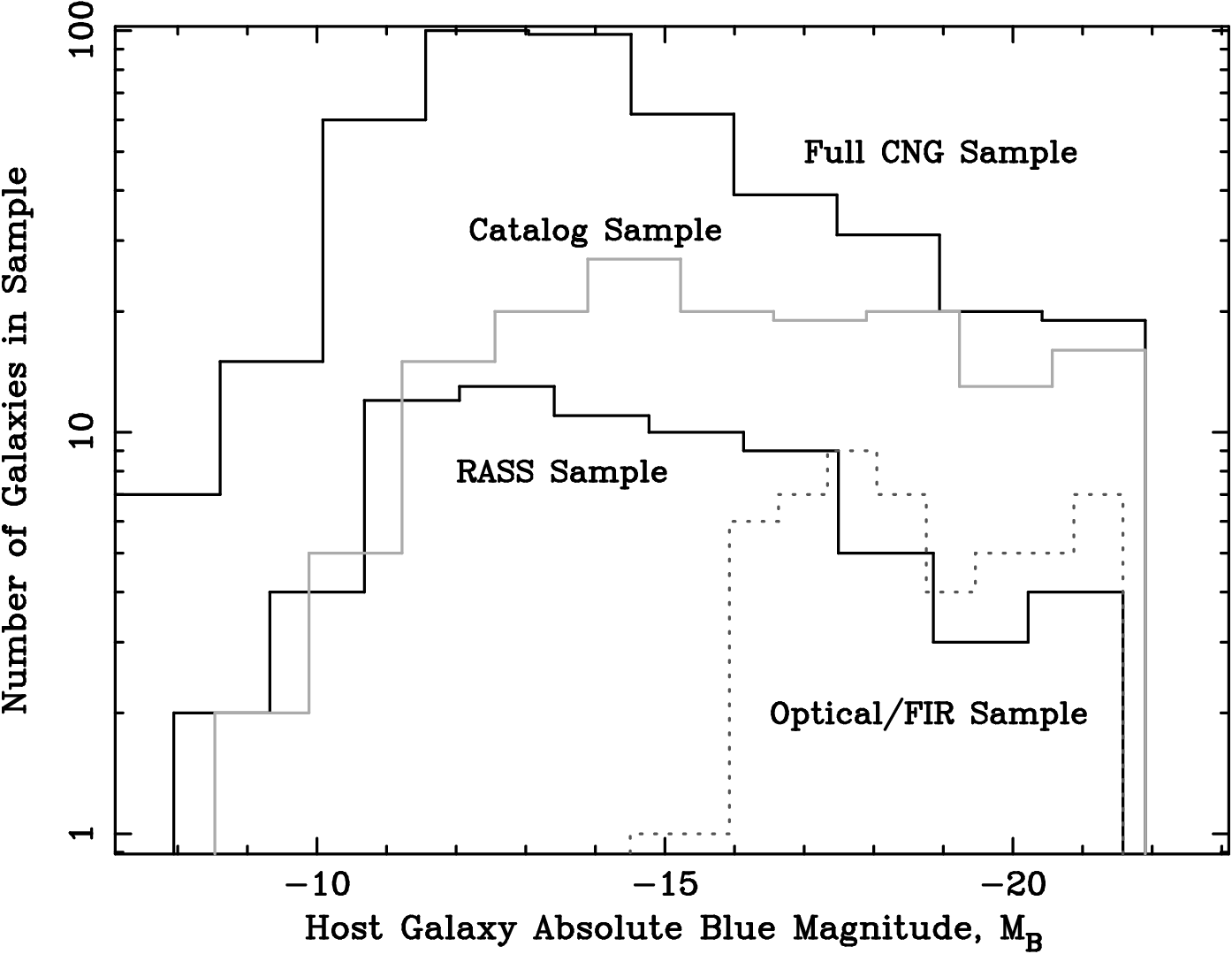}
\figcaption{Distribution of the number of galaxies in the full Karachentsev \etal\ (2004) catalog (uppermost curve) as a function of \mb\ and the distributions 
for three subsamples of the catalog as described in \S\S~2.1 (dotted historgram),
2.2 (lower heavy histogram), and 2.3 (light histogram).
\label{f:fig2}}
\end{center}

In all samples, only galaxies in the CNG with tabulated masses are 
considered. This is a subset of 313 of the 451 galaxies in the CNG. 
For this subset,
$\log(M/M_{\odot})=-0.39M_B+2.65$.
K04 define dwarf galaxies as those with $M_B>-17.0$~mag which 
corresponds to a mass $M<10^{9.3}$~\msun.
Galaxy mass is computed by K04 from H{\sc i} rotational velocity measurements
(inclination-corrected H{\sc i} line widths).
Consequently,
93 of the 138 galaxies omitted in the subset are gas-poor early-type galaxies
with revised Hubble type $T\le0$.
This leaves 5 giant and 10 dwarf early-tpe galaxies in the subset of 313 galaxies. 
The CNG sample of galaxies is shown as the uppermost histogram of Figure~2.
The histogram for the subset of galaxies with tabulated masses is similar.
As pointed out by K04, about 85\% of galaxies in the local volume are 
dwarf galaxies although these galaxies contribute only 4\% of the luminosity density.
From the CNG subset,
three subsamples are considered with different selection criteria and hence
different potential biases:

\subsection{
The optical {\sl and} FIR flux-limited subsample }
This subsample contains all galaxies within the Uppsala Galaxy Catalog 
of (northern hemisphere)
galaxies above the completeness limit of the UGC, $m_{ph}<14.5$~mag 
(Nilson 1973), with far-infrared (FIR) 
flux above the completeness limit of the {\sl Infrared Astronomy Satellite} (IRAS)
Point Source and Faint Source catalogs, $f_{\rm FIR} \sim 1.5$~Jy
(Beichman \etal\ 1988), and within 15 Mpc of the Milky Way. 
There are 49 galaxies that meet these criteria among the 313 CNG subset.
This subsample is biased toward the intrinsically optically-luminous galaxies with
high gas (dust) content and hence relatively high current star-formation rates. 
Figure~1 shows that the subsample is, indeed, comprised of relatively luminous 
and hence massive galaxies but that there are a few nearby low-luminosity 
(low-mass) objects included as well.

\subsection{
The nearby RASS subsample} \label{ss:rass}
This subsample contains all galaxies within 4.0~Mpc 
(a value based on expected count rates for ULXs,
 typical exposure times, and distances as 
tabulated by K04) located within 3\arcdeg\ of the centers of {\sl Rosat All Sky Survey} (RASS) fields. These fields are 6.4\arcdeg$\times$6.4\arcdeg\ regions
observed in scanning mode by the \ros/PSPC 
with typical exposures of $\sim$400~s in typical galaxy-size
regions of the sky. The RASS covers nearly
the entire sky but the search algorithm used 
to query the data archive\footnote{The High Energy Astrophysics Science Archive Research Center's on-line interface is http://heasarc.gsfc.nasa.gov/db-perl/W3Browse/w3browse.pl} uses circular 
search regions resulting in sampling 
only 70\% of the field's total area (using a 3\arcdeg\ search radius). There are
107 galaxies in the CNG subset within 4.0~Mpc and 72 or 67\% in the subsample.
This subsample contains the least bias of any of the subsamples. As shown in 
Figure~1, the distribution of blue luminosities for galaxies in this subsample
most closely matches the CNG distribution.

\begin{figure*}
\begin{center}
\includegraphics[width=0.6\columnwidth]{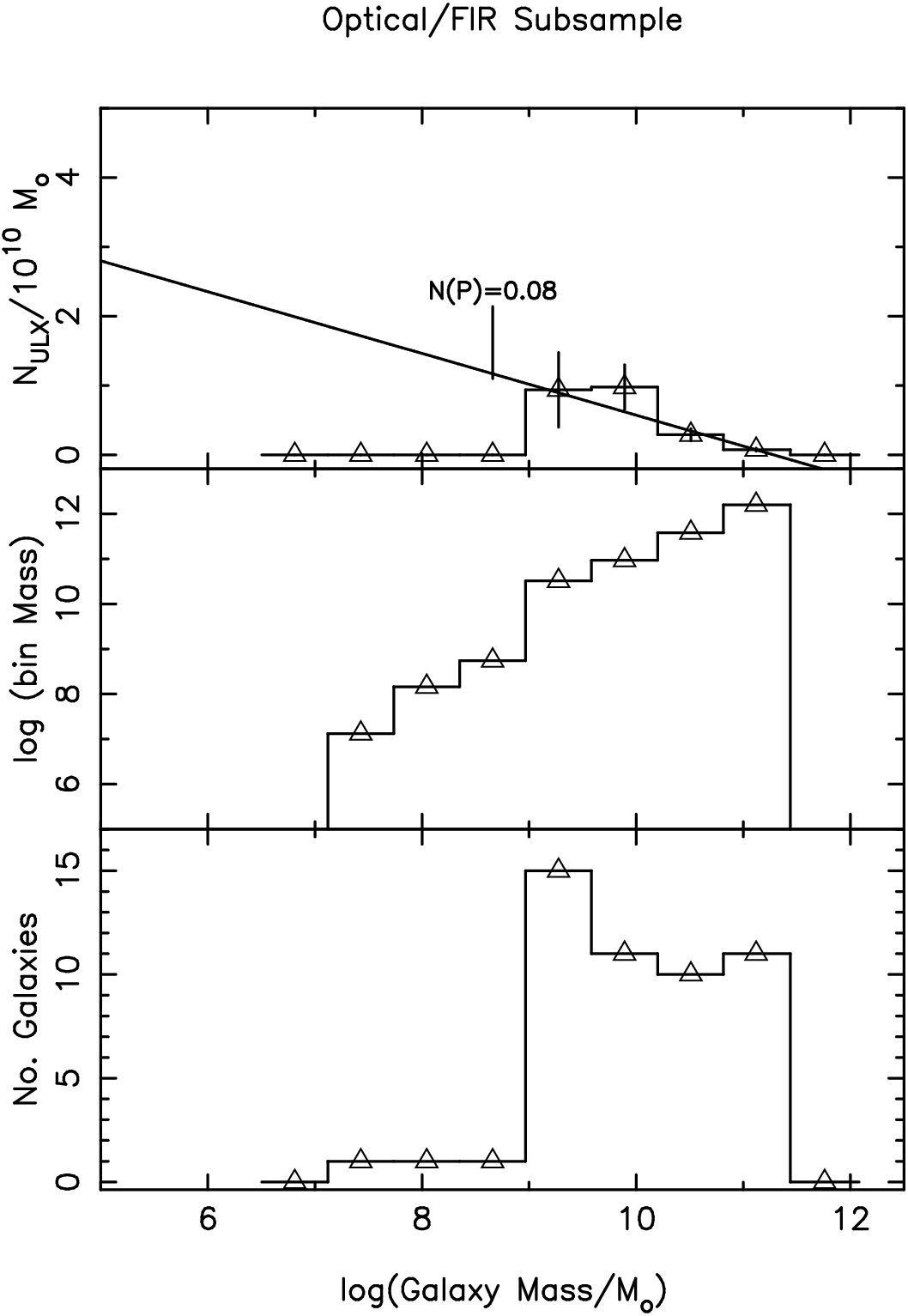} 
\includegraphics[width=0.6\columnwidth]{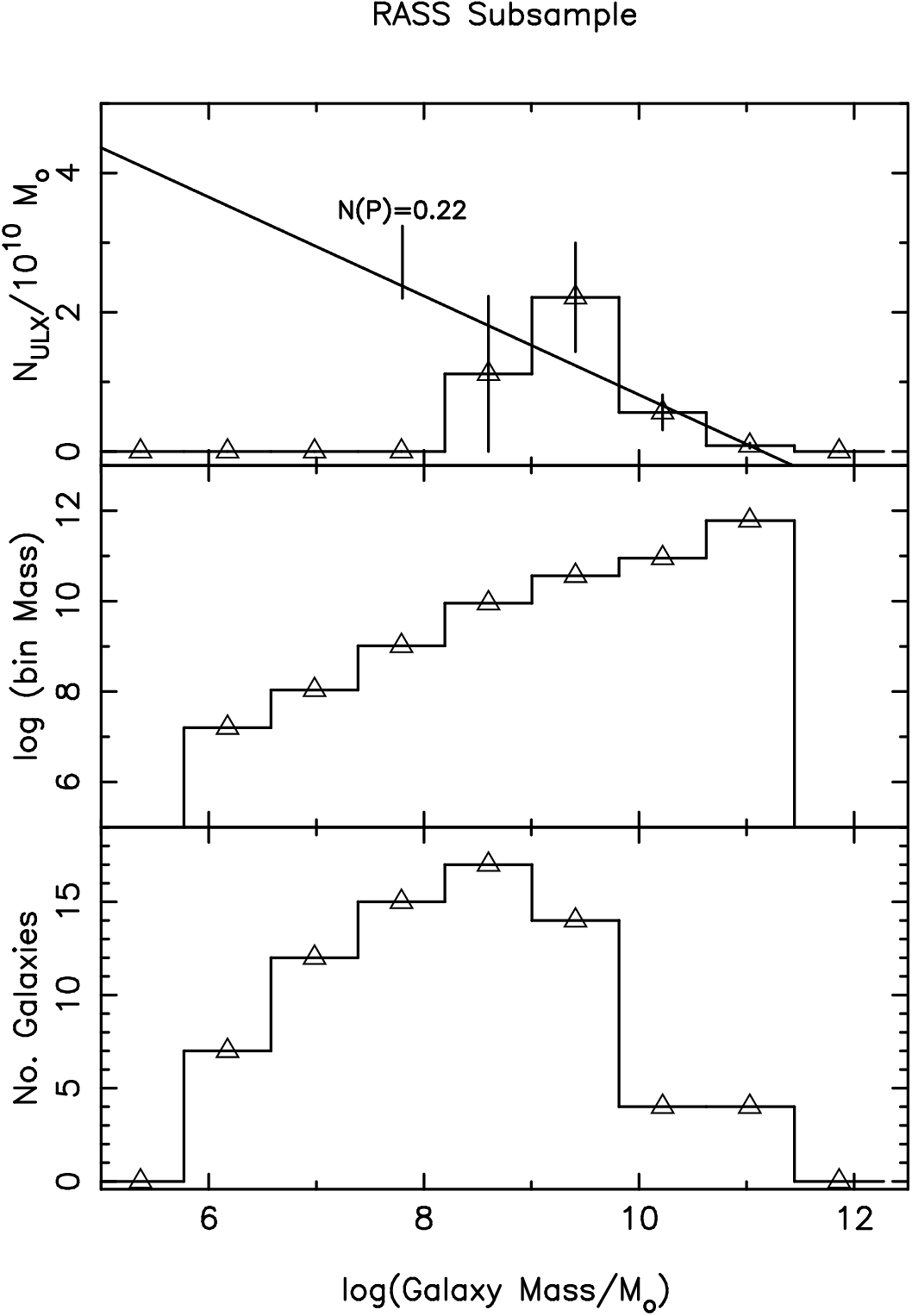} 
\includegraphics[width=0.6\columnwidth]{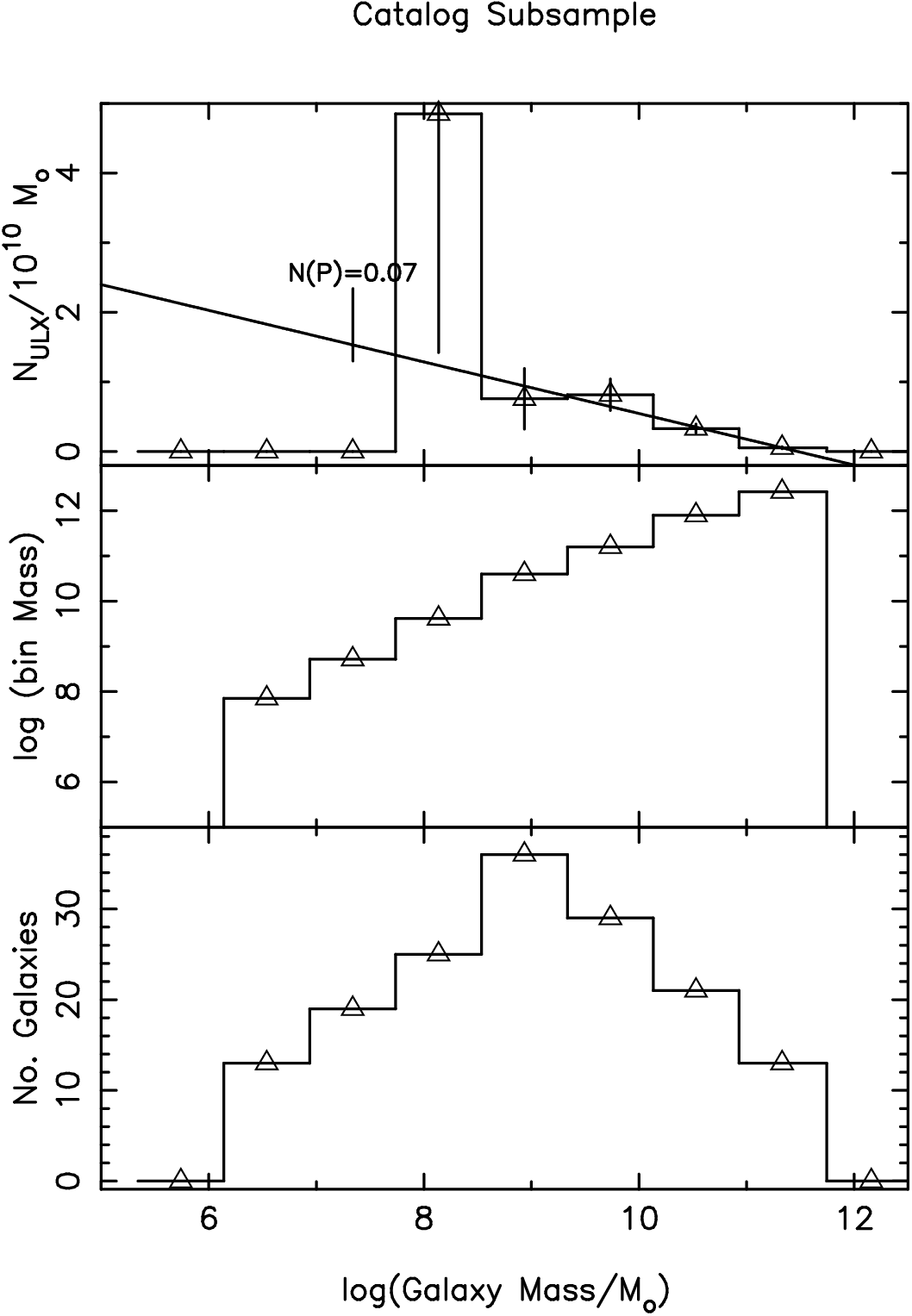}
\figcaption{Histograms of the specific frequency of candidate ULXs
(top panels), logarithm of the total mass in galaxies (middle panels),
and number of galaxies (lower panels) against the logarithm of the host galaxy mass
in solar mass units for three subsamples of galaxies tabulated in 
Karachentsev \etal\ (2004). From left to right, these subsamples are
described in \S\S~2.1, 2.2, and 2.3. In all graphs, the highest and lowest 
mass bin are added for clarity of the plots; they contain zero galaxies.
The lines shown in the upper panels are the best fit linear function to the 
data in bins with total mass exceeding $10^{10}$~\msun. In all
cases, these are the highest 4 bins containing galaxies. Labels in the upper
panels denote the number of ULX candidates predicted, $N(P)$, from extrapolation
of the linear function to the next lower mass bin and scaling by the relative mass
in that bin, $M_{10}$ (see text).
\label{f:fig3}}
\end{center}
\end{figure*}

\subsection{
The X-ray source catalog subsample} \label{ss:catalog}
This subsample contains all galaxies in the CNG subset that have been imaged in the
field of view of either \ros\ (PSPC or HRI), \cxo, or \xmm\ 
pointed observations of sufficient 
exposure to detect ULXs and that have had their ULX populations 
 identified through the
literature. This includes many CNG subset galaxies that were targets of 
pointed observations and,
in the case of \ros/PSPC with its 1\arcdeg\ radius FOV, several dwarf galaxies that 
were observed serendipitously. There are 155 galaxies in this subsample
including 69 serendipitously-observed galaxies (those with large offsets from
the observation's targeted aimpoint). This subsample should 
be considered an X-ray selected subsample since the majority of the 
galaxies were targets of pointed X-ray observations. It, like the Optical/FIR
subsample (see Figure~1), 
is biased towards the larger galaxies which are the preferential targets
of X-ray astronomy in general but it does contain a substantial number of 
dwarf galaxies.

\subsection{
Sample X-ray Analysis} \label{ss:xanalysis}

Both the literature and available X-ray imaging 
data (limited to \ros, \cxo, and \xmm)
were searched for candidate ULXs in this study. 
In most cases, whenever a thorough analysis of 
existing data was previously reported through the 
refereed literature, we did not perform an independent study.
In all cases, however, we did confirm that ULX
candidates were within the angular area of the purported 
host galaxy and not coincident with its nucleus.
For this,
an initial search was made for candidate ULXs within a circle of {\sl radius} $r=D_{25}$,
i.e., extending to twice the optical semi-major axis of the galaxy. 
This allows for potential astrometric misalignments between reported X-ray 
source positions and galaxy locations as tabulated in the CNG and
also for the large positional uncertainties that can occur in \ros\ and \xmm\ data
for weak sources because of their large point spread functions (PSFs).
If candidate ULXs were identifed, then the spatial coincidence was reconfirmed
 using an elliptical approximation of the \DTF\ isophote [with parameters taken 
from the Third Reference Catalog of Bright Galaxies (de~Vaucouleurs \etal\ 1991)], 
 a check of the host galaxy position against the 
 {\sl NASA/IPAC Extragalactic Database} (NED) position (and references therein)
 and an evaluation of the X-ray source positional uncertainty.
Sources coincident with host galaxy nuclei were rejected.
This resulted in the exclusion of 10 nuclear sources; all are the well-known
active galactic nuclei of massive galaxies.
Optical data (SDSS, DSS) were then visually inspected for 
 potential foreground stars and background active galaxy counterparts
which typically appear as bright and extended optical emitters, respectively. 
No such interlopers were identified using this test.
A check was also made to ensure the entire \DTF\ isophote was 
imaged in the X-ray observation. 
The several large galaxies not fully covered by \cxo's FOV have been fully sampled
by either \ros\ or \xmm\ observations; thus no ULX candidates were overlooked.

The source-detection algorithm developed by Tennant (2006) was applied to 
X-ray images whenever an analysis of the X-ray point source population 
for a galaxy was not published or was of limited utility. 
All the checks for spatial
coincidence with a host galaxy as outlined above were performed.
X-ray fluxes were estimated from observed count rates within a 3$\sigma$ source
radius (see Tennant 2006)
 using the {\sl Portable Interactive Multi-Mission Simulator} (PIMMS; Mukai 1993) 
and assuming an absorbed power law source spectrum.
The absorption was taken as the Galactic hydrogen column density along
the line of sight to the host galaxy as tabulated from the H{\sc i} map of 
Dickey \& Lockman (1990). 
The power-law index was assumed to be the average, $\Gamma=1.8$, 
obtained previously from
single-component spectral model fits to a large number of ULX candidates 
  by Swartz \etal\ (2004). 
X-ray luminosities were then estimated from the PIMMS fluxes 
and the distances tabulated by K04.
The choice of a single, fixed shape, model spectrum is appropriate
in the context of this study because only a small number of source photons 
($\lesssim$100) were typically
detected from candidate ULXs in those cases where thorough analysis 
was not previously reported.

Special mention must be made of the selection criterion of the RASS
subsample (\S~\ref{ss:rass}) and the subsequent analysis results. 
As mentioned above, 
the distance limit criterion was chosen 
based on the expected count rate for ULXs (using PIMMS)
and exposure times typical of the RASS and $\sim$10~counts needed for detection
(for each galaxy in this subsample,
the actual exposure time at the location of the galaxy
was determined from the exposure maps accompanying
the RASS X-ray event lists). 
Our source-finding analysis of the RASS data 
did not detect 10 of 18 ULX candidates previously described in
the literature (from pointed observations)
and expected to be visible in the RASS according to 
this simple perscription.
All non-detections are members of giant galaxies:
Six are buried within the strong diffuse emission of the starburst galaxy M82,
two are overlapping sources in RASS (though distinct in \cxo\ data) in NGC~4945,
one is a heavily-absorbed source in Circinus,
and one is a weak source in NGC~7793 that must have
faded compared to pointed \ros\ observations reported in
Read \& Pietsch (1999).
Thus, we attribute the non-detections to a combination of source 
confusion (aggravated by a rapidly-increasing off-axis PSF width;
see Hasinger \etal\ 1994),
high source column density 
(that decreases the  detectable flux within the soft \ros\ passband
from the PIMMS estimate assuming only a Galactic column density) 
and intrinsic source variability.
For these sources, we included the flux values obtained from
the literature.

\section{The Results}  

A total of 5 dwarf galaxies
($M<10^{9.3}$~\msun\ corresponding to $M_B>-17$~mag) 
were found to host a ULX in the present study. 
For comparison, there are a total of 35, 19, and 57 ULX candidates
in the three (overlapping) subsamples described, respectively, in \S\S~2.1--2.3.
Within these subsamples there are 10, 58, and 91 dwarf galaxies, respectively.
One new ULX candidate was discovered in a dwarf galaxy. This object was
detected in the RASS image of E059-01 (PGC~21199), an IB(s)m dwarf of mass
1.7$\times$10$^9$~\msun.

\subsection{The Specific ULX Frequency}  

Histograms displaying the number of ULX candidates per unit mass, 
 which we will refer to as the specific ULX frequency, 
against host galaxy mass were constructed for each 
 subsample.
The specific ULX frequency for the $i^{th}$ mass bin, $S^u_i$, is defined as 
$N_i/M_{10}^i\equiv\sum_j^{n_i} N_j/\sum_j^{n_i} M_{10}^j$ where 
$N_j$ is the number of ULX candidates detected in the $j^{th}$ galaxy, 
$M_{10}^j$ is that galaxy's mass in units of $10^{10}$~\msun, 
and the sum extends over all $n_i$ galaxies within the
mass interval comprising the $i^{th}$ mass bin.

Figure~2 displays the $S^u_i$ histograms for all three subsamples (upper panels)
the corresponding $\log(M_{10}^i)$ histograms (middle panels), 
and the number of galaxies, $N_i$,
contained in each mass bin (lower panels). 
The $S^u_i$ statistical errors are
$(N_i)^{1/2}M_{10}^i$.
There are no ULX candidates detected in galaxies with mass less than $10^{8.5}$~\msun. There is also very little total mass, $M^i_{10}$, in low-mass bins in the subsamples (middle panels, Figure~2).

We can estimate the number of ULXs expected in these lower-mass bins:
A simple linear function was fit to $S^u$ {\sl vs.} $\log(M/M_{\odot})$ 
over the mass bins with substantial total mass, $M^i_{10} \ge 1$,
in each subsample. There are 4 such bins, spanning roughly 3 decades in 
host galaxy mass, in each subsample.
From this, the number of ULXs predicted, $N_i(P)$ (where $P$ denotes predicted),
in the $i^{th}$ lower-mass bin is estimated by extrapolating this function
and weighting by the relative mass: 
$N_i(P)=f_i(S^u)M_{10}^i$ where $f_i(S^u)$ is the fitting function evaluated at
the $i^{th}$ mass bin. 
Values for $N_i(P)$ are given in the upper panels of Figure~2 along with the 
best-fitting function, $f_i(S^u)$.
$N_i(P)\ll 1$ in all cases because $M_{10}^i\ll 1$ for these lower-mass bins.
That is, there is simply too little mass in low-mass dwarf galaxies to expect to
find ULXs.

The fitting function used here was chosen to maximize
the number of ULXs predicted in lower-mass bins and hence provide a conservative
expectation value. 
There is no {\sl a priori} reason to expect the specific frequency of 
ULXs to increase with decreasing host galaxy mass (a more intuitive expectation is
that $S^u$ is independent of mass but it is clear from Figure~2 that $S^u$ equals 
a constant is a poor representation of the data). 

\subsection{The ULX Frequency \& The Star Formation Rate}  

We do, however,
 expect the number of ULX candidates to increase for galaxies with high 
current star formation rates (Grimm \etal\ 2003; Swartz \etal\ 2004; 
Colbert \etal\ 2004). In fact, from a sample of starburst galaxies, 
Grimm \etal\ (2003) find the number of X-ray sources above a luminosity, $L$,
depends linearly on the current star formation rate, $R_{SFR}$, as:
$N(>L) = 1.32 R_{SFR} (L_{39}^{-0.61}-21.0^{-0.61})$ where $R_{SFR}$ is measured
in units of \msun~yr$^{-1}$ and $L_{39}$ is the X-ray luminosity in units of
10$^{39}$~\ergl. Thus, we expect 1.1 ULXs per unit SFR.

The SFR for galaxies with a sizeable dust
opacity can be estimated from its FIR luminosity: 
$R_{SFR} = 4.5\times10^{-44} L_{FIR}$~\msun~yr$^{-1}$
(e.g., Kennicutt 1998). Here, $L_{FIR}$ is the 
FIR luminosity estimated from the IRAS 60$\mu$m and 100$\mu$m 
flux densities, $L_{FIR}/4\pi D^2=1.26\times10^{-11}(2.58S_{60}+S_{100})$~\ergl.
We have computed the star formation rate for
the galaxies in the optical and FIR flux-limited subsample (\S~2.1; many of the CNG galaxies are not included in the IRAS catalogs so we cannot make this computation
for the other subsamples) and constructed 
histograms of the distribution of ULXs with host galaxy mass as before but with
$M_{10}^j$ replaced with $R_{SFR}^j$ in the sums,
$\sum_j^{n_i} N_j/\sum_j^{n_i} R_{SFR}^j$, over the $n_i$ galaxies contained in
the $i^{th}$ mass bin. 

Figure~3 shows this distribution.
The ULX frequency normalized to unit SFR is roughly a constant for the bins
with substantial total mass. The best fit constant value is 2.3$\pm$0.7 which is 
about $2\sigma$ above the value predicted by Grimm \etal\ (2003).
A fit to a linear function is not a significant improvement according to an $F$-test.
Hence, the number of ULXs per unit SFR is similar for dwarf and giant galaxies
in this subsample. However, the SFR per unit mass is higher by about a factor of 3
for the dwarf galaxies compared to the giants which is consistent with the trend in
specific ULX frequency deduced above.

\begin{center}
\includegraphics[angle=-90,width=0.8\columnwidth]{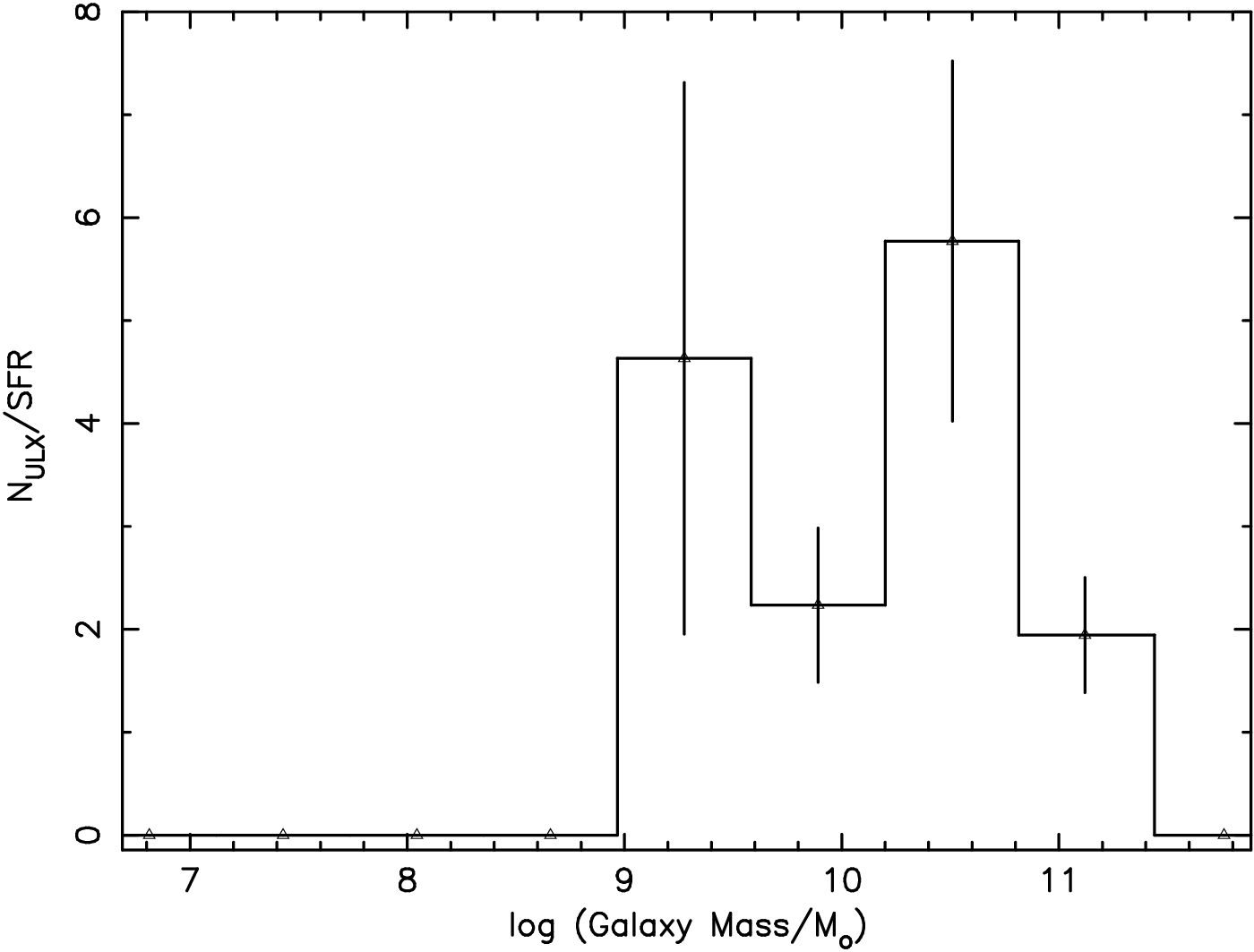}
\vspace{10pt}
\figcaption{Histogram of the number of candidate ULXs per unit
star formation rate (\msun~yr$^{-1}$) against the logarithm of the host galaxy mass (in solar mass units) for the galaxy subsample described in \S~2.1.
Total galaxy mass in each bin and number of galaxies per bin are
the same as in the lower two left-most panels of Figure~2.
\label{f:fig4}}
\end{center}

\subsection{Bias Assessment}  

There are three catagories of X-ray sources that may occur 
more or less frequently in dwarf compared to giant galaxies 
and thus have the potential to introduce biases in the present study: 
Contaminating background sources, 
transient ULXs, and nuclear sources.
Assuming the spatial
distribution of background sources
is uniform across the sky, the relative contributions of background sources to
the number of ULX candidates in each mass bin is proportional to the sum of the
 areas within the optical extents of the galaxies included in the bin. This is 
strictly true only for galaxies at a fixed distance because the number of 
background sources per unit area of sky increases with decreasing source flux
and the search for ULXs extends to lower fluxes for more distant galaxies.
However, the mean distance to dwarf galaxies in any of the subsamples (and to
the CNG subset with masses defined) is less than that to giant galaxies which
implies that potential contamination by background sources per unit area is higher
for the higher-mass bins considered. Furthermore, the total sky area covered by
galaxies in each mass bin increases towards higher-mass bins so, again, the 
potential contamination is higher for the higher-mass bins. 
Thus, any bias introduced by background sources applies in the opposite sense to
the trend determined here, namely, that the specific frequency of
ULXs increases with decreasing galaxy mass.

The probability of detecting transient sources increases with the rate
of observations sampling a particular target galaxy. 
The present study does not differentiate
between transient and non-transient ULX candidates but it can be surmised 
that the giant galaxies in the Local Volume
are more apt to have been observed multiple times over the 
course of the decades spanned by the \ros, \cxo, and \xmm\ missions than are
the dwarf galaxies. Thus, any bias introduced by ULX transience would 
favor an increase in the number of ULX candidates discovered
in giant over dwarf galaxies.

The nuclear regions of sample galaxies where not {\sl a priori} excluded from
our search for ULX candidates. 
However, X-ray 
luminous active galactic nuclei may hide nearby ULXs in their bright glow,
effectively imposing an exclusion region in the area of highest 
galactic surface mass density.
As the brightest AGN are associated with the most massive galaxies,
ULXs hidden by AGN brilliance could lead to an underestimate of the
ULX population in the higher-mass bins.
This is further exacerbated by the fact that 
the radial distribution of ULXs in general peaks toward the centers of galaxies
 (Swartz \etal\ 2004; Liu \etal\ 2006).
A total of 18 galaxies in the present study are listed as quasars or AGN
in the catalog of Veron-Cetty \& Veron (2006). Of these, 11 host X-ray bright
nuclear sources (all are giant galaxies). 
Fortunately, all have been observed with \cxo\ so that a careful
evaluation of the nuclear regions of these galaxies at high angular resolution
could be made. No evidence was found for hidden ULXs; thus, no bias is apparent
in the present results due to nuclear sources.

\section{Discussion}  

Ultraluminous X-ray sources occur in dwarf galaxies in the Local Volume;
at least down to galaxy masses of $\sim$3$\times$10$^{8}$~\msun. 
In fact, for the galaxies studied here, the specific frequency of ULXs in
low-mass galaxies is higher than in massive giants. 
This trend in the population of ULXs 
 is unanticipated. 

Still, ULXs in dwarf galaxies are rare, 
of order 1 ULX occurs per $10^{10}$~\msun\ in the galaxies studied here, 
which means only one
in 100 10$^8$~\msun\ dwarf galaxy is expected to host a ULX.
%
Five ULX candidates were identified in 118 dwarf galaxies
($M<10^{9.3}$~\msun\ corresponding to $M_B>-17$~mag) 
in the present study. 
Assuming the sample of galaxies was randomly selected, 
the Gehrels statistic (Gehrels 1986) gives a 90\% confidence 
that the true mean exceeds 2.4. 
Thus it is unlikely that our result is purely a statistical fluctuation. 
We detected no ULXs in galaxies with mass $<$10$^{8.5}$~\msun\ but,
if the trend continues to lower masses, then we expect $\le$0.22 ULXs to
have been detected in galaxies in our sample with mass $\sim$10$^{7.8}$~\msun\
which is consistent with finding 0.

It is worthwhile to consider what physical differences exist
between dwarf and giant galaxies that could give rise to the observed trend.
Past studies have concluded
that a strong correlation exists between star formation rate (SFR) 
and the number and luminosity density of ULXs
(Swartz \etal\ 2004; Liu \etal\ 2006; though see Ptak \& Colbert 2004).
For the one subsample purposefully chosen here to include FIR-luminous galaxies,
the specific current SFR in dwarfs is larger by a factor of 3 than in giants.
Thus, an increase in the specific ULX frequency in dwarf galaxies relative to 
giants would be expected at least for this subsample.

There are also, potentially, differences in the  stellar evolution and star formation 
processes in nearby dwarf galaxies compared to the giants that
may favor the formation of ULXs.
Dwarf galaxies tend to evolve more slowly than giants hence their metallicity at
the current epoch is systematically lower than that of giants
(Lee \etal\ 2006). 
Massive stars of low metallicity loose less mass through winds than do high-$Z$ 
stars and leave higher-mass compact objects following core collapse 
(Heger \etal\ 2003). Higher mass black hole remnants, in turn, can radiate at higher 
luminosities without violating the Eddington limit. Low-$Z$ donor stars may 
transfer more mass in Roche lobe overflow, and/or over a longer period of time,
 than do high-$Z$ donors with high wind mass loss rates. 
This would allow higher luminosity and longer-lived ULXs to arise in low-Z systems.

Dwarf galaxies form massive star clusters at a rate higher than expected for
their size.
Billet \etal\ (2002) explain this as due to a lack of shear in dwarf galaxies
that tends to fragment large molecular clouds 
and prevent formation of large stellar clusters in spiral galaxies.
There may be a connection 
between an excess of massive compact clusters and an excess of ULXs in dwarf galaxies. Both objects may be the end products of rapid collapse of molecular clouds with the ULXs formed from the most massive stars created in the collapse.
Gravitational coalescence of intermediate-mass protostars
 to form massive stars is observed  in
Galactic protoclusters (Peretto \etal\ 2007). The
cold initial conditions (low turbulent energy relative to gravitational energy)
needed for protostar merger are the same conditions envisioned by Billet \etal\ (2002) to explain the formation of massive clusters in dwarf galaxies.
These conditions may
 result from rapid compression by external processes caused by galactic-scale
interactions (e.g., Keto \etal\ 2005)~-- processes that are more effective in 
dwarfs because of their smaller size. 
Interestingly, Billet \etal\ (2002) conclude that the smallest galaxies 
 cannot sample the high end of the cloud mass spectrum and that
they will fall short of producing massive compact clusters. 
Perhaps a similar cutoff occurs in the production of ULXs and they are lacking 
 in very low-mass galaxies.

We have determined that ULXs occur in dwarf galaxies and that their specific 
frequency increases with decreasing host galaxy mass.
This suggests that there may be 
 special (though not unique) environmental conditions in dwarf galaxies that
 preferentially lead to the formation of ULXs.
This result should be addressed by theoretical models proposed to explain
 the ULX phenomenon.

\acknowledgements

This work was supported, in part, by Chandra Award GO6-7081A issued by the Chandra X-ray Observatory Center which is operated by the Smithsonian Astrophysical Observatory for and on behalf of NASA under contract NAS8-03060.

\end{document}